\documentclass[a4paper,11pt]{article}
\usepackage{pos}
\usepackage{xspace}
\usepackage{bm}
\renewcommand{\logo}{\relax}

\title{Magnetic winds in resistive compact binary discs}

\author*[a,b]{Marc Van den Bossche}
\author[a]{Geoffroy Lesur}
\author[a]{Guillaume Dubus}

\affiliation[a]{Univ. Grenoble Alpes, CNRS, IPAG, 38000 Grenoble, France}

\affiliation[b]{Leibniz-Institut für Astrophysik Potsdam (AIP),\\
An der Sternwarte 16, D-14482, Potsdam, Germany}

\emailAdd{mbossche@aip.de}

\abstract{Dwarf Novæ and low-mass X-ray binaries are eruptive binary systems comprised of a Roche-lobe overflowing solar-type star and an accreting compact object. Their recurrence time can be explained by a low-accreting phase, the quiescence, during which the angular momentum transport parameter is inferred to be $\alpha \approx 0.01$ by the Disc Instability Model. Non-magnetics mechanisms, such as spiral wave transport, only achieve angular momentum transport an order of magnitude too low, at best, because these discs are so thin in quiescence. During this phase, the Magneto-rotational Instability is known to be suppressed by the increased resistivity of the weakly ionised plasma. Studying these thin magnetised discs is a numerical challenge because of the wide range of scales to be resolved. 
Thanks to the new GPU-accelerated code Idefix, we produce global 3D MHD simulations of a very thin disc $(H/R = 0.01)$ for the first time. We explore the possibility that an MHD wind arises and increases accretion in low magnetic Reynolds number $(\mathrm{Rm}\approx100)$ and realistic plasma parameter ($\beta \approx 1000$) regimes. We observe that the MRI is only quenched in the resistive disc bulk but survives in the disc atmosphere. This drives strong accretion and wind launching. We quantify the efficiency of the resulting wind and measure its global effect on the disc. We explore the effect of the initial disc magnetisation and compare the accretion/ejection regime with and without resistivity.
}

\FullConference{87th Fujihara Seminar: The 50th Anniversary Workshop of the Disk Instability Model in Compact Binary Stars (DIM50TH2025)\\
22-26 September 2025\\
Tomakomai, Japan\\}

\newcommand{\Mspy}{~\ensuremath{\mathrm{M}_\odot/\mathrm{yr}}\xspace}

\begin{document}
\maketitle

\section{Introduction}

Compact binaries such as dwarf novæ and X-Ray Binaries regularly go into bright outbursts; their recurrence cycle is now well described by the Disc Instability Model (DIM), initially proposed by \cite{osaki_accretion_1974}, and further developed by \cite{mineshige_disk-instability_1983,meyer_outbursts_1984}. See, for example, \cite{hameury_review_2020} for a review of the model.

The difference in luminosity between these systems during the outburst and quiescence phases is due to a dramatic change in the accretion regime. During the hot and bright outburst phase, accretion onto the central object is strong, with accretion rate of $10^{-9}\Mspy$ to $10^{-8}\Mspy$. During this phase, hydrogen in the disc is fully ionised and the plasma is well coupled to any present magnetic field. This allows the magneto-rotational instability (MRI, \cite{velikhov_stability_1959,chandrasekhar_hydrodynamic_1961,balbus_powerful_1991}) to drive strong turbulence inside the disc. This turbulence behaves as an effective viscosity, allowing for the transport of angular momentum and accretion. This effective viscosity is quantified by the $\alpha$ parameter introduced by \cite{shakura_black_1973}. Provided large-scale convection takes place in the disc \cite{hirose_convection_2014}, this instability produces accretion levels of $\alpha = 0.1$ that match estimates from observation \cite{smak_dn_1999,king_accretion_2007,kotko_viscosity_2012}.

On the other hand, the dim and cold quiescence phase remains poorly understood. First, because it is much dimmer, it proves challenging to obtain direct observational constraints from these systems. Second, the temperature of the disc during quiescence estimated using the DIM is close to 3,000~K. This corresponds to a very thin disc, with extreme aspect ratio close to $H/R = 10^{-3}$, where $H$ is the hydrostatic scale-height of the disc and $R$ its radius. Such thin discs are hard to study with simulations as they require very high resolution to resolve structures even smaller than the disc scale height.

Still, we know from X-Rays observations that there still is accretion during this phase \cite{wheatley_xrays_2003,pandel_xray_2005}. The corresponding $\alpha$ values can be estimated from the recurrence times of outbursts and \cite{cannizzo_outburst_1988,cannizzo_kepler_2012} find $\alpha =0.02 - 0.04$ for the quiescence phase.

Moreover, \cite{gammie_origin_1998} showed that during the quiescence phase, the disc was very poorly ionised. As a consequence, the MRI can not drive the turbulence necessary to induce the observed accretion, as the disc is too resistive and the disc's matter and magnetic field are only poorly coupled. 

Since then, understanding the mechanism driving the accretion during the quiescence phase has been a very fundamental open question of the DIM.

The first possibility that was explored is spiral-shock-driven accretion \cite{savonije_tidal_1983,savonije_tidally_1994}. These spiral waves, excited by the tidal interaction of the close companion star, produce strong shocks that contribute to angular momentum transport. This is a compelling mechanism for the quiescence phase as it does not rely on the coupling of the magnetic field to the disc material. The theoretical models predicted that the efficiency of this mechanism should drop with the disc temperature. Recently, we showed in \cite{vdb_spiral_2023} with the first high-resolution simulations of quiescent spiral-driven-accretion, that this mechanism indeed drives $\alpha < 10^{-4}$ in realistically cold quiescent discs. We noted, however, that the linear theory in these analytical works breaks down in this regime because the shocks become very sharp and strong.

Scepi \textit{et al.} \cite{scepi_impact_2018, scepi_turbulent_2018} proposed two possibilities. In their shearing-box simulations, they first observed that the hot part of the cold branch of the hysteresis cycle may be in a regime where the disc is hot enough for the MRI to survive. Second, in the presence of a large-scale vertical magnetic field, magnetic winds may be launched from the disc \cite{blandford_hydromagnetic_1982}. This mechanism is especially relevant for quiescence as magnetic winds drive accretion by exerting a torque on the disc surface, but do not produce the viscous heating associated with usual viscous-like accretion-driving effects. Using their previous results from shearing-box simulations, \cite{scepi_magnetic_2019} added the effects of magnetic wind in the DIM and were able to reproduce dwarf-novæ-like light curves with a dipolar magnetic field.

In this talk, we presented the first 3D global simulations of quiescent dwarf novæ discs to study the magnetic wind hypothesis in a global model. The numerical setup is briefly described in the next section. We shortly list our results in the last section. A more in-depth presentation of the methods and results of this work can be found in an upcoming article (Van den Bossche\textit{ et al. } in prep).

\section{Numerical methods}

\subsection{Equations}

In this work, I use the finite-volume Idefix code \citep{lesur_idefix_2023}. This code uses the Kokkos library for performance portability \cite{CarterEdwards_kokkos_2014}. All the simulations presented here were run on GPU (both Nvidia V100 on Jean Zay, IDRIS and AMD Mi250 on Adastra, CINES). This allowed us to probe realistic quiescence regime for the first time with a global MHD model. The code solves the MHD equation in spherical coordinates $(r,\theta,\phi)$.

\begin{align}
    &\partial_t \rho + \nabla \cdot ( \rho \bm{v} ) = 0,\\
    &\partial_t(\rho \bm{v}) + \nabla  \cdot ( \rho \bm{v}\bm{v}^T ) = - \nabla p - \rho \nabla \Psi + \frac{1}{4\pi}\left ( \nabla \times \bm{B}\right ) \times \bm{B},\\
    &\partial_t\bm{B} = \nabla \times \left ( \bm{v} \times \bm{B} \right ) + \nabla \times \left ( \eta \nabla \times \bm{B} \right)
\end{align}
with $\rho$ the fluid density, $\bm{v}$ the fluid velocity, $p$ its pressure, $\Psi$ the gravitational potential, $\bm{B}$ the magnetic field, and $\eta$ the Ohmic resistivity.

We close these equations with a barotropic equation of state:
\begin{equation}
    p = c_\text{s}^2(\rho) \rho,
\end{equation}
with
\begin{equation}
    c_s (\rho) = c_\mathrm{d} + \frac{1}{2} \left [1 - \tanh \left ( \frac{\log \frac{\rho}{\rho_\mathrm{c}}}{\Delta}\right ) \right ]\left ( c_\mathrm{c} - c_\mathrm{d} \right ),
    \label{eq:EoS}
\end{equation}

This equation of state corresponds to a two-temperature fluid. The high density material, with $\rho \gg \rho_\mathrm{c}$, is cold (with sound speed $c_\mathrm{d}$) and the low density material, with $\rho \ll \rho_\mathrm{c}$, is hot ($c_\mathrm{d}< c_\mathrm{c}$). $\Delta$ is the width over which the regime transition occurs. The critical density $\rho_\mathrm{c}$ is chosen such that the disc mid-plane density is several orders of magnitude above it. The disc atmosphere is made out of the hot material; this is to model irradiation heating of the low-density corona and also insure vertical stability of the disc and atmosphere.

The temperature of the disc is chosen such that the aspect ratio of the disc is $H/R = 0.01$ at the inner radius. As this is still slightly hotter than an actual quiescent disc, we introduce a temperature rescaling in the self-consistent computation of the local Ohmic resistivity. We do so in such a way that the magnetic Reynolds number $\mathrm{Rm} = \frac{\Omega^2 H}{\eta}$, with $\Omega$ the local orbital angular frequency, has the value of an actual quiescent disc. In our simulations we consider a very resistive disc, with $\mathrm{Rm}<50$ in the inner regions, and $\mathrm{Rm}<1000$ everywhere, that is below the MRI threshold.

\subsection{Initial conditions}

For the initial condition of our simulations, we take the analytical solution to the hydrostatic equilibrium of the disc, with surface density $\Sigma \propto R$, as predicted by the DIM for the quiescence phase. We use 

\begin{equation}
    \rho (R,z) = \rho_0 \left (\frac{R}{r_0} \right)^{\frac{3}{2}} \exp \left [ \frac{GM_\mathrm{WD}}{c_\mathrm{s}^2} \left (\frac{1}{\sqrt{R^2+  z^2}} -  \frac{1}{R}\right )  \right ],
\end{equation}
where $M_\text{WD}$ is the mass of the white dwarf,  $ R = r \sin \theta$ and $z = r \cos \theta$. We pre-truncate the disc at its expected outer truncation radius (like in \cite{vdb_spiral_2023}). We first evolve this state for ten binary orbits without magnetic field in an axisymmetrical 2.5D setup with no tidal potential (with potential $\Psi_\mathrm{init}$). We do this in order to reach numerical equilibrium of the initial non-magnetic state. We then restart the simulation in the full 3D domain and introduce at this time the magnetic field and full gravitational potential $\Psi_\mathrm{full}$.

One has 
\begin{equation}
        \Psi_\mathrm{init} (\bm{r})= - \frac{GM_\text{WD}}{\vert \vert \bm{r}\vert \vert},
\end{equation}
\begin{equation}
        \Psi_\mathrm{full} (\bm{r})= - \frac{GM_\text{WD}}{\vert \vert \bm{r}\vert \vert} - \frac{GM_\text{s}}{\vert\vert\bm{r} - \bm{a}\vert\vert } +\frac{GM_\text{s}}{a^3}\bm{a}\cdot\bm{r},
\end{equation}

where $G$ is the gravitational constant, $M_\text{s}$ is the mass of the companion star, and $\bm{a}$ is the binary separation. We use $q = \frac{M_\text{s}}{M_\text{WD}} = 0.4$ to avoid any resonance instability occurring for $q < 0.3$.

The magnetic field introduced at this point is slightly slanted, similarly to \cite{zanni_mhd_2007}, to reduce the duration of the initial transient state. This field is initialised by the potential vector

\begin{equation}
    A_r = A_\theta = 0, \quad     A_\phi = \frac{4}{3} B_0 \left (\frac{R}{r_0}\right )^{m+2} r_0^2 \frac{\kappa^{5/4}}{(\kappa^2 + z^2/R^2)^{5/8}/R}
\end{equation}

with $\kappa$ the length scale over which the field is bent and $r_0$ the inner radius of the domain. We set its amplitude $B_0$ such that $\beta_0 = \frac{8 \pi P}{B_0^2}$ is constant in the entire disc. We produce simulations  with $\beta_0 = 10^3$ and $\beta_0 = 10^4$.

We also perform a small parameter exploration, in addition to the different values of $\beta_0$. We perform simulations with and without Ohmic resistivity and a reference simulation with no magnetic field at all.

\begin{figure}
    \centering
    \includegraphics[width=\linewidth]{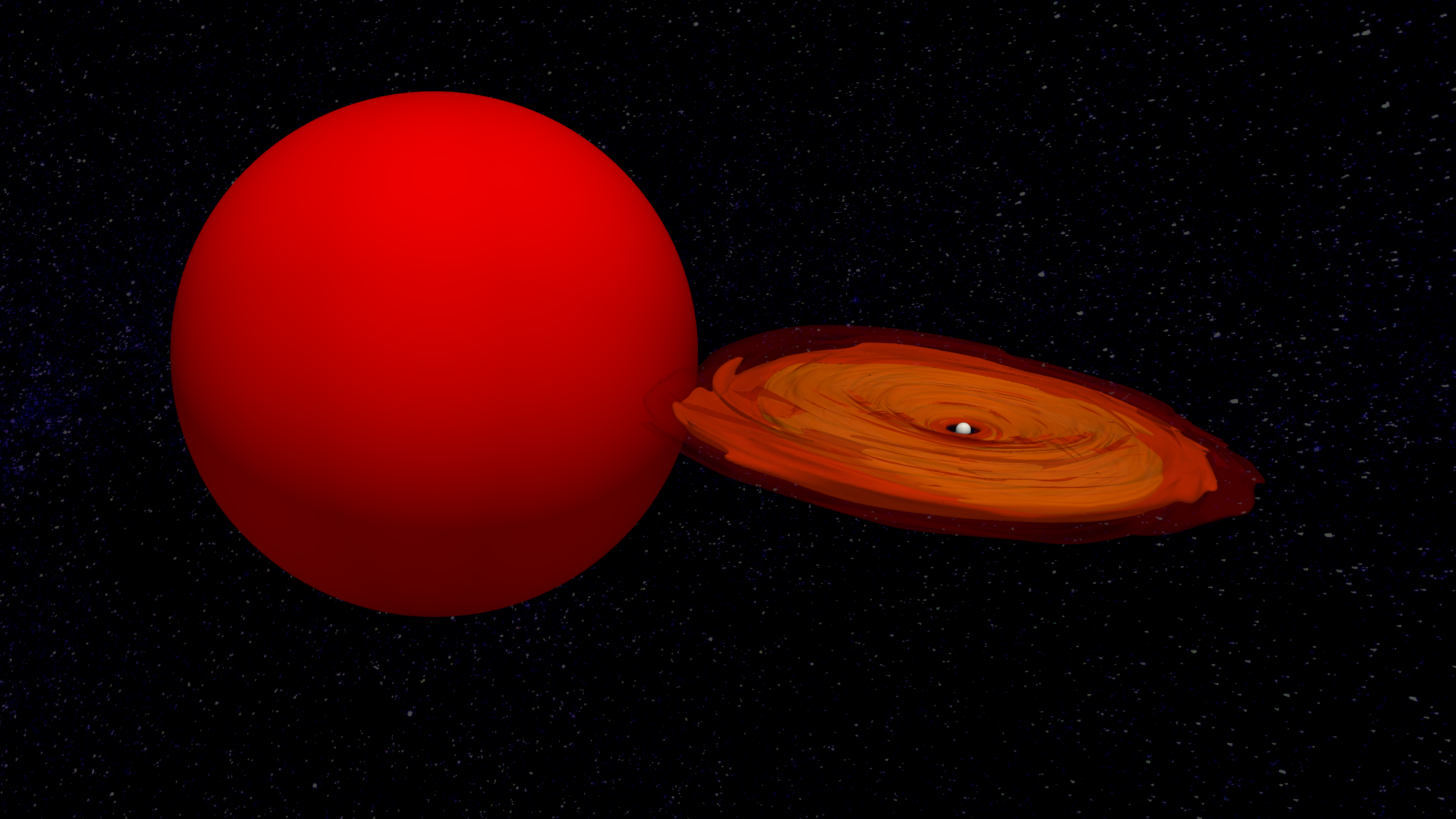}
    \caption{3D rendering of a snapshot of the simulations of this work. The coloured surfaces are density iso-contours of the disc. The red sphere is a representation of the companion star, the white sphere is a representation of the white dwarf. These spheres are to scale and are added for the render only. So is the background.}
    \label{fig:illusration}
\end{figure}

\section{Results overview}

\noindent  Figure \ref{fig:illusration} illustrates the thin disc simulations of this work.

\vspace{2em}

\noindent Our conclusions are

\begin{enumerate}
    \item Even from a very highly resistive disc, magnetic winds are launched from the $\beta \approx 1 $ regions, as long as the initial magnetic field is strong enough.
    \item The MRI is quenched in the disc bulk, but does survive at the surface. The Reynold magnetic number is no longer the relevant criterion to predict the existence of the MRI in this regime.
    \item Accretion is strongly enhanced by the presence of a magnetic field, but it is not dominated by the magnetic wind.
    \item The disc is strongly puffed up due to the additional contribution of the magnetic pressure in the vertical equilibrium.
    \item The produced magnetic wind is strongly non-top-bottom symmetric, with one side of the disc having a stronger wind ejection rate than the other.
\end{enumerate}

A more in-depth discussion of these results will be published in an upcoming article (Van den Bossche\textit{ et al. } in prep).

\newpage
\bibliographystyle{JHEP}
\bibliography{bib}

\end{document}